\documentclass{ws-procs9x6}

\begin{document}

\title{PHOTON REGENERATION PLANS\footnote{Talk presented at the 6th International Workshop on the Identification of Dark Matter (IDM 2006), Island of Rhodes, Greece, 11--16th September, 2006.}
}

\author{A. RINGWALD$^\dagger$}

\address{Deutsches Elektronen-Synchrotron DESY,\\ 
Notkestra\ss e 85, D-22607 Hamburg, Germany\\
$^\dagger$E-mail: andreas.ringwald@desy.de\\
www.desy.de/\~{}ringwald}

\begin{abstract}
Precision
experiments exploiting low-energy photons may yield 
information on particle physics complementary to experiments at
high-energy colliders, in particular on
new very light and very weakly interacting
particles, predicted in many extensions of the standard model. 
Such particles may be produced by 
laser photons send along a transverse magnetic field. The laser polarization
experiment PVLAS may have seen the first indirect signal 
of such particles by observing an anomalously large rotation of the 
polarization plane of photons after the passage through a magnetic field. This
can be interpreted as evidence for photon disappearance due to particle production.  
There are a number of experimental proposals to test independently the particle interpretation
of PVLAS. Many of them are based on the search for photon reappearance 
or regeneration, i.e. for ``light shining through a wall''.  
At DESY, the Axion-Like Particle Search (ALPS) collaboration
is currently setting up such an experiment. 
\end{abstract}

%\keywords{}

\bodymatter

\section{Introduction}

The standard model of particle physics is phenomenologically extremely successful. 
However, there are a number of hints which point to the possibility that there is new
physics beyond it. Proposed extensions of the standard model aim at a solution of
important problems such as the unification of all forces, including
gravity, or at an explanation of the absence of $CP$ violation in strong
interactions, among many others.   

Many attempts to embedd the standard model into a more general, unified framework, 
notably the ones based on string theory, predict, apart from new very heavy, $m\gg 100$~GeV, 
particles also a number of new very light, $m\ll 1$~eV particles, which are very weakly 
coupled to ordinary matter. Prominent candidates for such particles go under the 
names axions (arising in the course of a solution of the strong $CP$ problem), 
dilatons, and millicharged particles. 

\section{Polarization Experiments}

\begin{figure}
\begin{center}
\psfig{file=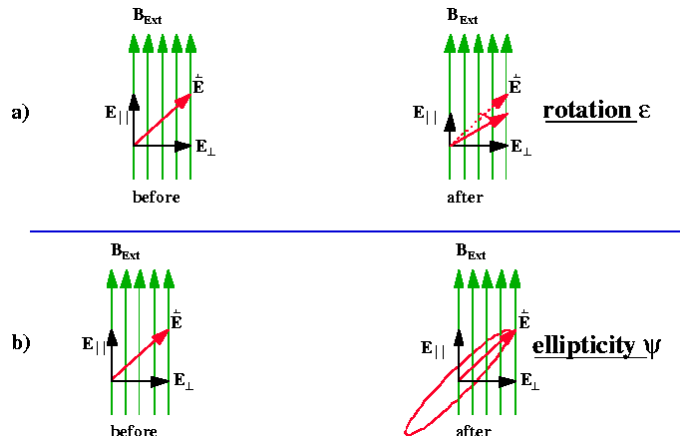,width=3.5in}
\end{center}
\caption[]{Possible changes of the polarization state of initially linearly polarized 
photons after the passage through a magnetic field (adapted from Ref.~\cite{Brandi:2000ty}).}
\label{fig:pol}
\end{figure}

Laboratory experiments to search for such particles may be based on the possibility 
to produce them by shining laser photons along a transverse magnetic field. 
Searches for a change of the polarization state of initially linearly polarized 
photons after the passage through the magnetic field, in particular for a possible
rotation (dichroism) and ellipticity (birefringence) (cf. Fig.~\ref{fig:pol}), 
are particularly sensitive to new light particles~\cite{Maiani:1986md,Gies:2006ca}. 

%%%%%%%%%%%%%%%%%
%%%%%%%%%%%%%%%%

In a pioneering laser polarization experiment, the BFRT collaboration established an 
upper limit both on a possible vacuum magnetic (VM) dichroism
and birefringence~\cite{Cameron:1993mr}. 
Recently, however, the PVLAS collaboration reported the observation
of a VM dichroism~\cite{Zavattini:2005tm}. Moreover, preliminary data seem to indicate 
also evidence for a 
non-vanishing VM birefringence. These observations have led to a number of theoretical
and experimental activities, since the magnitude of the reported signals exceeds the
standard model expectations by far.    

%%%%%%%%%%%%%%%%%
\begin{figure}
\begin{center}
\psfig{file=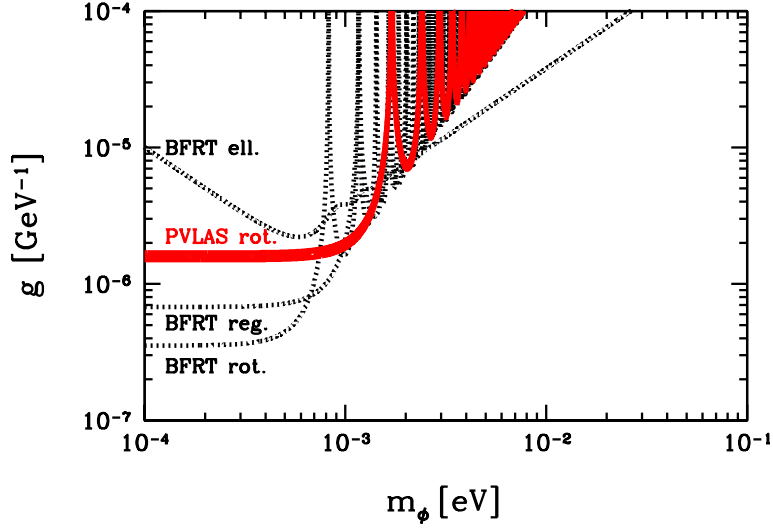,bbllx=25,bblly=226,bburx=580,bbury=604,width=4.in}
\end{center}
\caption[]{ALP interpretation of BFRT and PVLAS data: two photon coupling $g$ versus mass $m_\phi$. 
The 95\,\% confidence level upper limits from BFRT data~\cite{Cameron:1993mr}  
on polarization (rotation and ellipticity data)  
and photon regeneration are displayed as dotted lines. 
The preferred values corresponding to the anomalous rotation signal 
observed by PVLAS~\cite{Zavattini:2005tm} are shown as a thick solid line.}
\label{fig:alp_int}
\end{figure}
%%%%%%%%%%%%%%%%

\begin{figure}
\begin{center}
\psfig{file=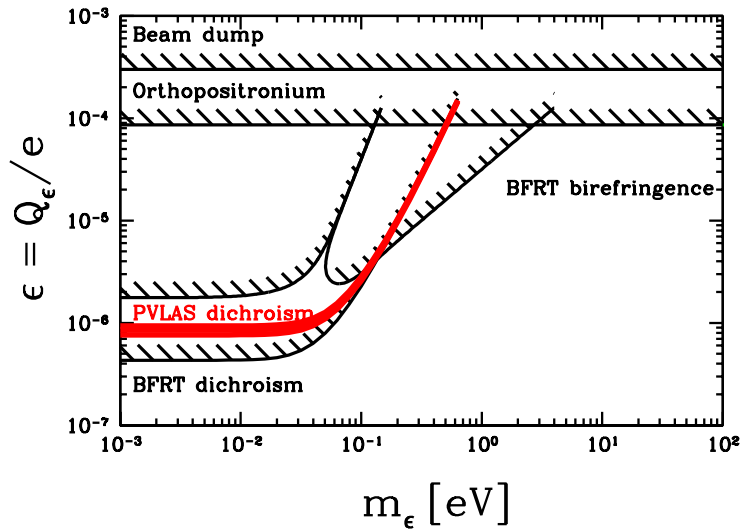,bbllx=25,bblly=226,bburx=580,bbury=604,width=4.in}
\end{center}
\caption[]{MCP interpretation of BFRT and PVLAS data: fractional electric charge 
$\epsilon = Q_\epsilon/e$ versus mass $m_\epsilon$. 
The preferred values corresponding to the anomalous rotation signal 
observed by PVLAS~\cite{Zavattini:2005tm} are shown as a thick solid line.
}
\label{fig:mcp_int}
\end{figure}

Two viable particle physics explanations of the reported signals have been 
proposed: the real and virtual production\\ 
(i) of a neutral spin-0 (axion-like) 
particle (ALP) $\phi$~\cite{Maiani:1986md}
 with mass $m_\phi$ and a coupling to two photons via 
\begin{equation}
{\mathcal  L}^{(+)}_{\rm{int}}
  =-\frac{1}{4}g\phi^{(+)}F_{\mu\nu}F^{\mu\nu}
  =\frac{1}{2}g\phi^{(+)}(\vec{E}^{2}-\vec{B}^{2}),
\end{equation}
or
\begin{equation}
{\mathcal L}^{(-)}_{\rm{int}}
  =-\frac{1}{4}g\phi^{(-)}F_{\mu\nu}\widetilde{F}^{\mu\nu}
  =g\phi^{(-)}(\vec{E}\cdot\vec{B}),
\end{equation}
depending on its parity, denoted by the superscript $(\pm)$, 
or\\ 
(ii) of a pair of millicharged, $Q_\epsilon =\epsilon e$,
particles (MCP) $\epsilon^+ \epsilon^-$~\cite{Gies:2006ca} 
with mass $m_\epsilon$, coupling to photons in the
usual way via the minimal substitution $\partial_\mu \to D_\mu \equiv \partial_\mu -{\rm i}\epsilon
e A_\mu$ in the Lagrangian.\\  
Indeed, as apparent from Fig.~\ref{fig:alp_int}, the 
rotation observed by PVLAS can be reconciled with the non-observation
of a rotation and ellipticity by BFRT, if there is an ALP 
with a mass $m_\phi\sim$~meV and a coupling
$g\sim 10^{-6}$~GeV$^{-1}$~\cite{Zavattini:2005tm}. 
Alternatively, both experimental results are compatible with the 
existence of an MCP with $m_\epsilon\sim 0.1$~eV and $\epsilon \sim 10^{-6}$~\cite{Gies:2006ca} 
(cf. Fig.~\ref{fig:mcp_int}). 
These parameter values, however, seem to be in serious conflict with 
astrophysical bounds, arising for example from energy loss considerations 
of stars~\cite{Raffelt:1996,Davidson:2000hf}. 
They may be evaded if the production of ALPs or MCPs is
suppressed in astrophysical plasmas~\cite{Masso:2005ym,Jaeckel:2006id}, 
as realized, for example, in 
extensions of the standard model involving extra U(1) gauge bosons which
kinetically mix with our familiar hypercharge U(1)~\cite{Masso:2006gc}, 
such as in some realistic string compactifications~\cite{Abel:2006qt}.     
 
%%%%%%%%%%%%%%%%%
%%%%%%%%%%%%%%%%

It is very comforting that a number of laboratory-based low-energy
tests of the ALP and MCP interpretation of the PVLAS anomaly are
currently set up and expected to yield decisive results within the
upcoming year. For example, in addition to PVLAS, the BMV~\cite{Rizzo:Patras} 
and Q\&A~\cite{Chen:2006cd} collaborations 
will run further polarization experiments with different experimental parameter 
values which finally may lead to a  discrimination between the ALP and the MCP 
hypothesis~\cite{Ahlers:2006iz}.

%%%%%%%%%%%%%%%%%
\begin{figure}
\begin{center}
\psfig{file=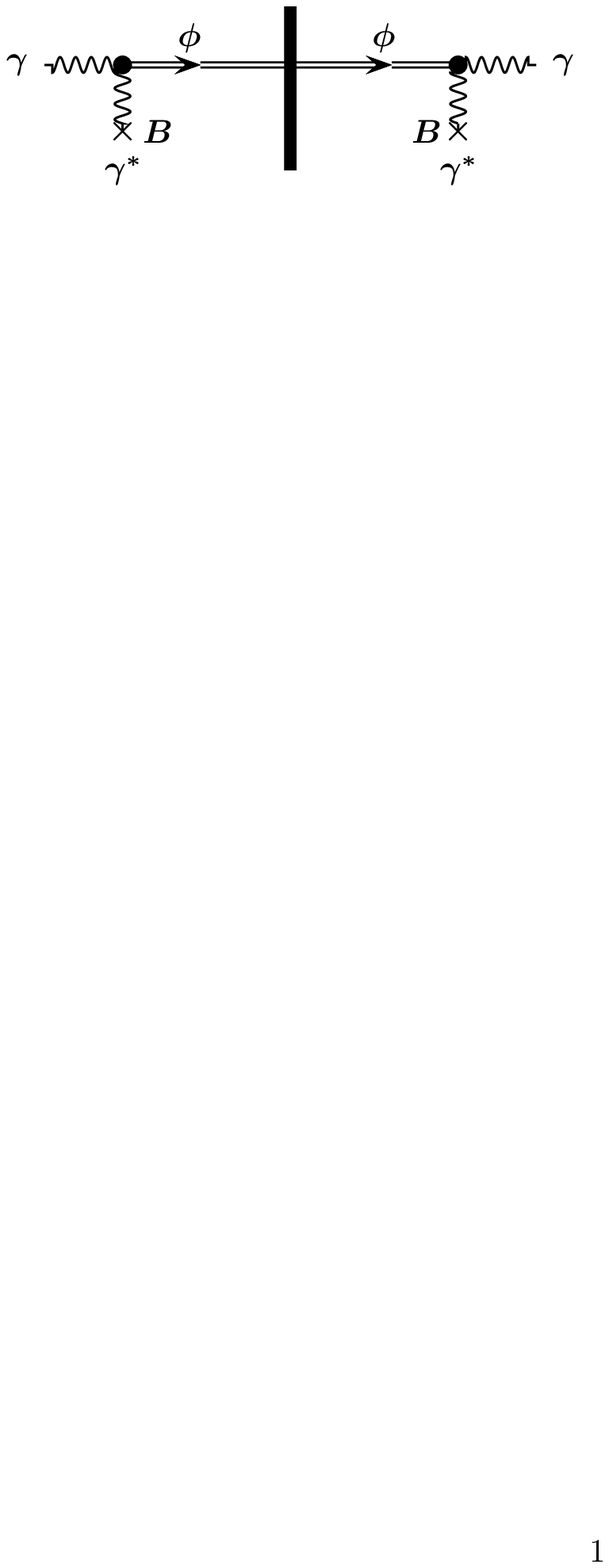,bbllx=86pt,bblly=637pt,bburx=298pt,bbury=707pt,width=3.5in}
\end{center}
\caption[]{Schematic view of ALP production through photon conversion 
in a magnetic field (left), subsequent travel through a wall, and final 
detection through photon regeneration (right). }
\label{fig:reg}
\end{figure}
%%%%%%%%%%%%%%%%

\section{Regeneration Experiments}

The ALP interpretation of the PVLAS signal will crucially be tested by photon
regeneration (sometimes called ``light shining through walls'')
experiments~\cite{Sikivie:1983ip,Anselm:1986gz,VanBibber:1987rq,Ringwald:2003ns}, presently 
under construction or serious 
consideration~\cite{Rizzo:Patras,ALPS,Baker:Patras,Cantatore:Patras,Pugnat:2005nk} 
(cf. Table~\ref{tab:exp}).  
In these experiments (cf. Fig.~\ref{fig:reg}), a photon beam is directed across
a magnetic field, where a fraction of them turns into ALPs. The ALP
beam can then propagate freely through a wall or another obstruction
without being absorbed, and finally another magnetic field located on
the other side of the wall can transform some of these ALPs into
photons --- seemingly regenerating these photons out of nothing.
A pioneering photon regeneration experiment has been done also by the
BFRT collaboration~\cite{Cameron:1993mr}. 
No signal has been found and the corresponding upper limit on 
$g$ vs. $m_\phi$ is included in Fig.~\ref{fig:alp_int}.

%%%%%%%%%%%%%%%%%%%
\begin{table}
\tbl{Experimental parameters of proposed photon regeneration experiments: 
magnetic fields $B_i$ and their length $\ell_i$ on production ($i=1$) and 
regeneration ($i=2$) side (cf. Fig.~\ref{fig:reg}); 
and the corresponding probability $P_{\gamma\phi\gamma}$, 
with $g\sim 2\times 10^{-6}$~GeV$^{-1}$ in the PVLAS preferred region 
(cf. Fig.~\ref{fig:alp_int}).}
{\begin{tabular}{|l|l|c||l||}
\hline
Name & Laboratory   & Magnets & $P_{\gamma\phi\gamma}$ for $g\sim 2\times 10^{-6}$~GeV$^{-1}$ \\
\hline
{\bf ALPS}~\cite{ALPS}&DESY/D     & $B_1=B_2=5$~T & 
\\
 &  	      & $\ell_1=\ell_2=4.21$~m &  $\sim 10^{-19}$  \\
\hline
{\bf BMV}~\cite{Rizzo:Patras} &LULI/F 	    & $B_1=B_2=11$~T &
\\
 &        & $\ell_1=\ell_2=0.25$~m &  $\sim 10^{-21}$  \\
\hline 
{\bf LIPSS}~\cite{Baker:Patras}&Jlab/USA    & $B_1=B_2=1.7$~T & 
\\
 &    & $\ell_1=\ell_2=1$~m & $\sim 10^{-23.5}$  \\
\hline
 &    & $B_1=5$~T & 
\\
{\bf PVLAS}~\cite{Cantatore:Patras} & Legnaro/I     & $\ell_1=1$~m &   $\sim 10^{-23}$
\\
	 &  			& $B_2=2.2$~T & \\
&            &$\ell_2=0.5$~m  &\\
\hline
{\bf ---}~\cite{Pugnat:2005nk}&CERN/CH    & $B_1=B_2=11$~T & 
\\
 &        & $\ell_1=\ell_2=7$~m &  $\sim 10^{-17}$ \\
\hline
\end{tabular}}
\label{tab:exp}
\end{table}
%%%%%%%%%%%%%%%%%%%%%%%%%%%%%%%%%

Clearly, crucial experimental parameters for such an experiment are
the magnetic fields $B_i$ and their length $\ell_i$ on the production ($i=1$) and
the regeneration ($i=2$) side of the apparatus. Indeed,   
the conversion probability for the process $\gamma\to\phi\to\gamma$  
is given by
\begin{eqnarray}
\nonumber
P_{\gamma\to\phi\to\gamma}&=&P_{\gamma\to\phi}(B_1,\ell_1,q_1)\,P_{\phi\to\gamma}(B_2,\ell_2,q_2)\,,
\\ \label{regpropab}
P_{\gamma\to\phi}(B,\ell,q) &=&P_{\phi\to\gamma}(B,\ell,q) = 
\frac{1}{4} \left(g\,B\,\ell\right)^2\,F(q\ell)\,,
\end{eqnarray}
where $F(q\ell )\leq 1$ is a form factor which equals unity, if the photons 
and the ALPs act coherently over the whole length of the magnet, i.e. for   
\begin{equation}
q\ell = \left|\frac{2\,(n_\gamma -1)\,\omega^2-m_\phi^2}{2\omega }\right| \ell\ll 1 ,
\end{equation}
$\omega$ being the photon energy and $n_\gamma$ being the refraction index of eventual buffer gas 
in the beam pipe~\cite{vanBibber:1988ge}. 
At small momentum transfer, $q\ell\ll 1$, therefore, the regeneration probability scales as
$(B_1\ell_1 B_2\ell_2)^4$. Correspondingly, strong and long dipole magnets from high-energy 
storage rings (e.g. HERA~\cite{Ringwald:2003ns} or LHC) 
are particularly suited for a photon regeneration experiment, 
as can be seen in Table~\ref{tab:exp}. 

%%%%%%%%%%%%%%%%%
\begin{figure}
\begin{center}
\psfig{file=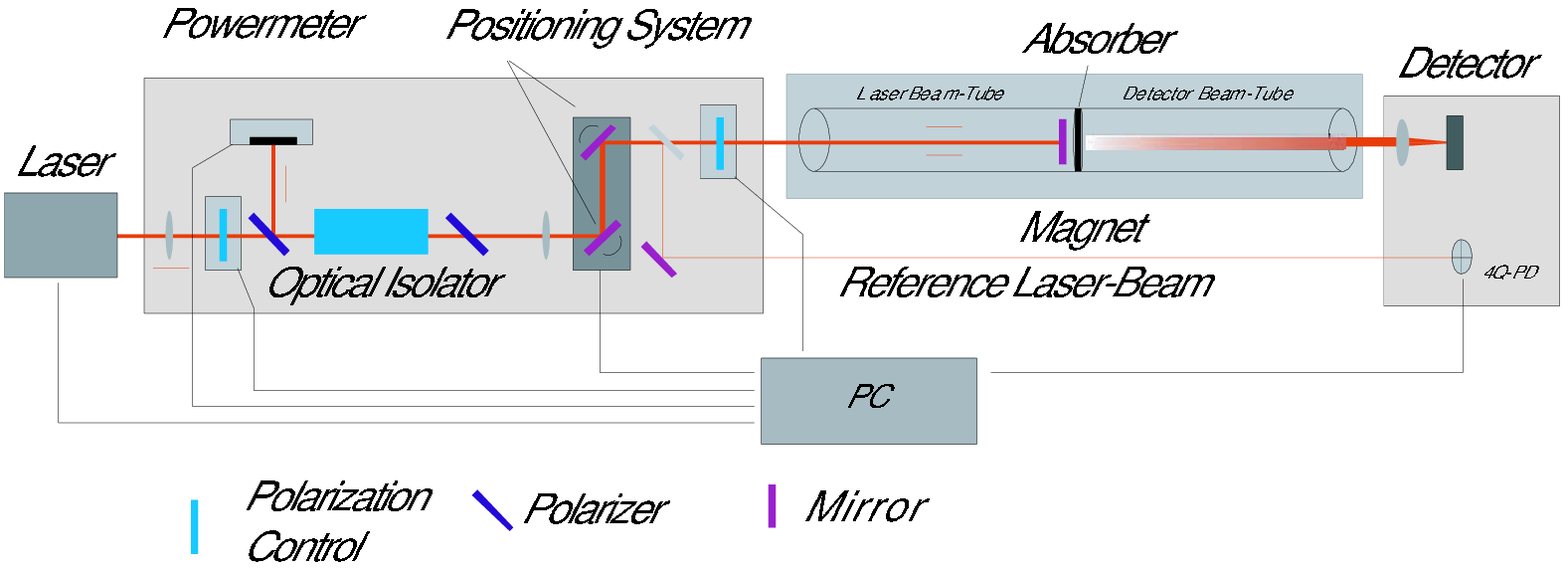,width=4.1in}
\end{center}
\caption[]{Schematic view of the experimental setup of the ALPS experiment~\cite{ALPS}.}
\label{fig:ALPS}
\end{figure}
%%%%%%%%%%%%%%%%

\begin{figure}
\begin{center}
\psfig{file=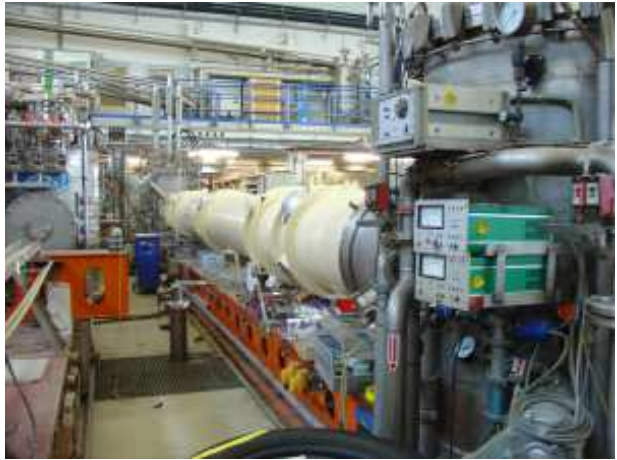,width=3.2in}
\end{center}
\caption[]{HERA dipole magnet at DESY's magnet test stand. It will be 
exploited for the ALPS experiment~\cite{ALPS}.}
\label{fig:heradip}
\end{figure}

\begin{figure}
\begin{center}
\psfig{file=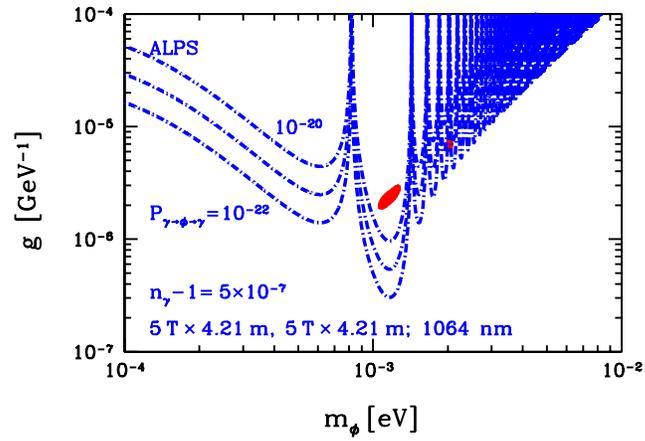,bbllx=25,bblly=226,bburx=580,bbury=604,width=3.3in}
\end{center}
\caption[]{Isocontours of the probability $P_{\gamma\to\phi\to\gamma}$, 
Eq.~(\ref{regpropab}), for the parameters of ALPS experiment: $B_1=B_2=5$~T, 
$\ell_1=\ell_2=4.21$~m, exploiting an infrared photon beam ($\omega =1.17$~eV)
in buffer gas with a refractive index $n_\gamma -1=5\times 10^{-7}$ (from Ref.~\cite{ALPS}).}
\label{fig:alps_sens}
\end{figure}

Importantly, one may optimize the sensitivity in 
certain mass regions by essentially tuning $q$ towards small values by adjusting $n_\gamma$, 
i.e. by varying the gas pressure in the magnetic field regions~\cite{vanBibber:1988ge}.  
This effect will be heavily exploited at the Axion-Like Particle Search (ALPS) 
experiment~\cite{ALPS} at DESY (cf. Fig.~\ref{fig:ALPS}), 
where the photon beam of a high-power, $\sim 200$~W, 
infrared laser,
corresponding to an initial flux of $\sim 10^{21}$ photons/s, 
will be sent along the $\sim 5$~T, $\sim 8.8$~m transverse magnetic field of a superconducting 
HERA dipole magnet (cf. Fig.~\ref{fig:heradip}). 
Filling in a buffer gas with a refractive index $n_\gamma -1 \sim 10^{-7}$, 
one will have the maximum sensitivity in the PVLAS preferred parameter 
region (cf. Fig.~\ref{fig:alps_sens}). ALPS,  a  collaboration of  DESY, 
the  Laserzentrum  Hannover  and the  Sternwarte  Bergedorf,  
has been approved  in  principle by the 
DESY directorate and is planning to take data in summer 2007.  

%%%%%%%%%%%%%%%%%
%%%%%%%%%%%%%%%%
%%%%%%%%%%%%%%%%%
%%%%%%%%%%%%%%%%

\section{Conclusions}

The evidence for a vacuum magnetic dichroism found by PVLAS has triggered a lot
of theoretical and experimental activities. Fortunately, in the upcoming year
a number of decisive laboratory based tests both of the axion-like and  
millicharged particle interpretation will be done. 
In particular, the planned photon regeneration experiments (cf. Table~\ref{tab:exp})
will firmly establish or exclude the axion-like particle interpretation. 
In summary, even in the LHC era, 
small experiments might have a big impact on particle physics!

\bibliographystyle{ws-procs9x6}
\bibliography{ws-pro-sample}

\end{document}